\documentclass{optica-article}

\journal{opticajournal} 

\articletype{Research Article}

\usepackage{graphicx}
\usepackage{siunitx}
\usepackage{amsmath}
\usepackage{nicefrac}
\usepackage{bm}
\usepackage{amsmath}
\usepackage{appendix}
\usepackage{titlesec}
\usepackage{xcolor}
\usepackage{amsmath}
\newcommand{\revised}[1]{\textcolor{black}{#1}}      

\graphicspath{ {./images/} }

\usepackage{url}

\usepackage{microtype}

\newcommand\blfootnote[1]{%
  \begingroup
  \renewcommand\thefootnote{}\footnote{#1}%
  \addtocounter{footnote}{-1}%
  \endgroup
}

\begin{document}

\blfootnote{Accepted for publication in Optics Express. The published version, when available, will be at \url{https://doi.org/10.1364/OE.610722}}

\title{Efficient Modeling of 3D-Printed Freeform Waveguides with a Transformation Optics-Based Beam Propagation Method}

\author{Sina Foroutan-Barenji,\authormark{} Jonas Krimmer,\authormark{} Stefan Singer,\authormark{} Christian Koos,\authormark{} and Wolfgang Freude\authormark{*}}

\address{\authormark{}Institute of Photonics and Quantum Electronics (IPQ), Karlsruhe Institute of Technology (KIT), Engesserstr. 5, 76131 Karlsruhe, Germany\\}

\email{\authormark{*}w.freude@kit.edu} 


\begin{abstract*}
An efficient numerical method is introduced for simulating wave propagation in 3D-printed freeform waveguides, referred to as photonic wire bonds (PWBs). This approach integrates transformation optics (TO) with curvilinear meshing to map the original 3D freeform PWB structure onto a straight waveguide, which possesses anisotropic material characteristics. The resulting waveguide is then analyzed using an anisotropic full-vectorial beam propagation method (BPM). To validate the performance and effectiveness of the proposed TO-BPM algorithm, a comparison is made with the finite-difference time-domain (FDTD) technique. Our approach \revised{reduces the simulation time by more than two orders of magnitude, depending on the degree of trajectory bending,} in combination with a \revised{mean absolute percentage error (MAPE) of \qty{3.9}{\percent}} in power transmission simulation relative to the FDTD.
\end{abstract*}

\section{Introduction}
Photonic integrated circuits (PICs) have emerged as essential building blocks for a broad spectrum of applications, enabling advancements across various technological domains. These include high-speed communication systems~\cite{dong2014silicon}, cutting-edge artificial intelligence frameworks~\cite{shen2017deep}, ultra-fast signal processing techniques~\cite{drayss2023non}, precision optical metrology~\cite{trocha2018ultrafast}, as well as innovative solutions in life sciences and biophotonics~\cite{kohler2021biophotonic}. The ability to integrate multiple photonic functions on a single chip has revolutionized these fields, providing unparalleled performance, miniaturization, and energy efficiency. From a technological perspective, photonic integrated circuits (PICs) primarily utilize planar designs, which are fabricated through well-established microfabrication techniques, including layer deposition and two-dimensional patterning via high-resolution electron-beam or deep-UV lithography ~\cite{ito2000pushing}. In recent years, these traditional fabrication approaches have been complemented by two-photon lithography, which allows for the in-situ creation of three-dimensional freeform structures, thereby significantly improving the capabilities of planar PICs~\cite{xu2021hybrid,NesicLAM,xu20233d}. An especially intriguing method involves the use of 3D-printed freeform waveguides, referred to as photonic wire bonds (PWBs)~\cite{lindenmann2012photonic}. These waveguides facilitate the interconnection of optical chips, such as those found in high-performance photonic multi-chip modules~\cite{xu2021hybrid,NesicLAM}. A crucial feature of the PWB concept is its capability to address inherent alignment errors in the optical chips. This is achieved by adjusting the trajectory and geometry of the PWB to match the precise locations of the coupling facets. \revised{With trajectory we mean the curve $(x_0, y_0, z_0)$ through the waveguide center.} These positions are determined solely during the fabrication process and must be converted into a low-loss PWB design within seconds before printing. Currently, this is achieved using computationally efficient heuristic algorithms for PWB design, which select a smooth path to link the relevant optical interfaces~\cite{lindenmann2012photonic,negredo2018fast,nesic2022transformation}. Although this method is quick and produces satisfactory outcomes, it is \revised{purely geometric and does not solve Maxwell’s equations. Therefore it is} unable to predict the PWB loss. In contrast, solvers based on first-principle Maxwell's equations, such as those using the finite-difference time-domain (FDTD) method, have been investigated for PWB modeling~\cite{nesic2022transformation}. However, even when adapted to the specific case of moderately bent planar waveguides through the use of transformation optics (TO), these solvers remain too slow~\cite{nesic2022transformation}. The beam-propagation method (BPM) has been used for the efficient simulation of wave propagation in both isotropic and anisotropic media. However, due to its assumption of paraxial propagation, the technique encounters challenges when dealing with freeform bent structures~\cite{feit1978light}. Various BPM algorithms have been investigated to enhance the speed of electromagnetic simulations for integrated optical waveguides. Xu \textit{et al.}~\cite{xu1994full} \revised{and Selleri \textit{et al.}~\cite{selleri2000}} introduced \revised{full-vectorial BPM approaches} suitable for planar waveguides with anisotropic materials. Lu \textit{et al.}~\cite{lu2002beam} proposed a method using a scalar BPM combined with a curvilinear coordinate system to model bent slab waveguides. Kawamura \textit{et al.}~\cite{kawamura2020design} applied a full-vectorial BPM along with coordinate transformation to model tapered planar waveguides. An efficient Maxwell-based 3D freeform waveguide modeling and design tool is hence still lacking.
\begin{figure}[b!]
    \centering
    \includegraphics[width=\textwidth]{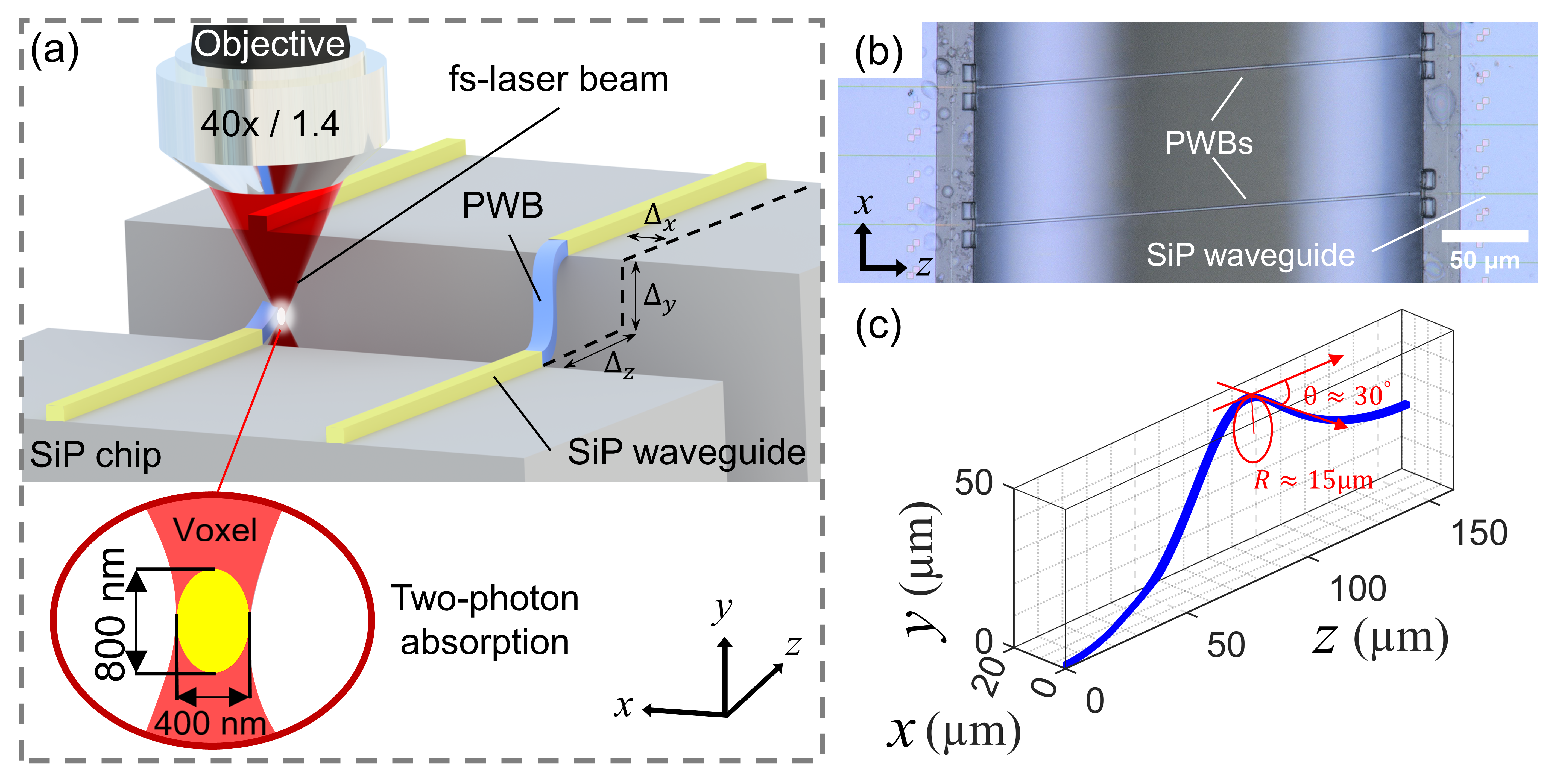}
    \caption{Design and fabrication of photonic wire bonds (PWB). A liquid resin is solidified by two-photon absorption (TPA) of a focused fs-laser beam. (a) Two SiP waveguides on two separate chips have a lateral offset $\Delta_x$,  a vertical offset  $\Delta_y$, and a longitudinal separation $\Delta_z$. Along a trajectory  with smooth transitions based on splines, a  PWB is 3D-printed \emph{in-situ} for interconnecting two SiP waveguides. (b) \revised{Top-view micrograph ($x$--$z$ plane) of exemplary fabricated freeform PWBs interconnecting silicon-photonic (SiP) waveguides}, with offsets $\Delta_x = \qty{10}{\micro\meter}$, $\Delta_y = \qty{30}{\micro\meter}$ and a longitudinal separation $\Delta_z = \qty{270}{\micro\meter}$. \revised{The PWBs are the short bright segments in the center, bridging the two straight SiP waveguides on either side.} The white scale bar is $\qty{50}{\um}$ long. (c) Side view of the 3D trajectory of an exemplary freeform PWB with a specifically small curvature radius $R_{\mathrm{min}}\approx\qty{15}{\um}$ and a steepest-slope angle $\theta\approx30^\circ$. This sharp bend does not allow the computation with  a conventional paraxial BPM method.}
    \label{fig:design}
\end{figure}

In our previous conference contribution~\cite{foroutan2024efficient}, we presented preliminary results on an TO-BPM algorithm. \revised{The present work is a substantially extended version, providing the full theoretical framework, the rigorous formulation of the curvilinear second-order mapping, a convergence study, and the treatment of both polarizations.} TO-BPM is a highly efficient algorithm, designed specifically for simulating wave propagation in 3D-printed freeform PWBs. The method relies on a full-vectorial BPM and combines it with TO. In particular, TO is employed alongside curvilinear meshing to convert the initial 3D freeform PWB into a straight waveguide with anisotropic material characteristics, which is subsequently analyzed using BPM. \revised{In contrast to the planar transformation-optics approach of Nesic \textit{et al.}~\cite{nesic2022transformation}, the present method handles the non-diagonal transformed dielectric tensors of non-planar three-dimensional freeform trajectories, and follows the curved waveguide boundaries directly. With a 3D freeform (non-planar) trajectory we mean a waveguide whose center is a space curve bending in all three spatial dimensions, in contrast to a planar waveguide such as rib or slab structures whose axes remains within a plane. In addition, rather than assigning constant material properties to large blocks, as previously done in ~\cite{nesic2022transformation}, our method assigns the varying transformed material to each node.} We validate the effectiveness of our TO-BPM algorithm by comparing its performance to the FDTD method. For practically relevant PWB designs with moderate bending, our method achieves an up to \revised{300}-fold decrease in computation time while achieving a \revised{mean absolute percentage error (MAPE) of \qty{3.9}{\percent}} in power transmission simulations compared to the FDTD benchmark values. To the best of our knowledge, our work represents the first implementation of a full-vectorial BPM that can handle 3D freeform waveguides having non-plane trajectories. We anticipate that this method could play a crucial role in enabling highly efficient real-time routing and designing of PWB structures in high-throughput optical assembly processes.
\section{Methods}
As depicted in Fig.~\ref{fig:design}(a), PWBs establish optical connections between different PIC chips~\cite{lindenmann2012photonic}. The PWB relaxes alignment tolerances for photonic chips (Fig.~\ref{fig:design}(b)) and fibers, enabling placement without the need for expensive active alignment\revised{~\cite{lindenmann2015connecting}. Horizontal and vertical offsets of up to several tens of micrometer can be tolerated as well as longitudinal displacements of a few hundred micrometer. As long as the bending radius stays above about \qty{50}{\micro\meter}, the loss is reasonably small.} The PWB is fabricated by focusing a femtosecond beam into a liquid \revised{photoresist} using a high NA immersion objective, triggering localized polymerization (i.e., solidification) via two-photon absorption. \revised{For the simulations, the material properties of the photoresist VanCore~B (Vanguard Automation GmbH) are used. The material is optimized for printing micro-optical structures with high shape fidelity, and has a refractive index \(n = 1.53\) at \(\lambda_0 = \qty{1.55}{\micro\meter}\).} The PWB is defined by moving the focal spot laterally and axially\revised{~\cite{billah2018hybrid,nesic2019photonic}}. In the end, the residual liquid ink is removed during the development process, resulting in a printed PWB (Fig.~\ref{fig:design}(c)). The trajectory of the PWB is designed for smooth transitions based on splines\revised{~\cite{nesic2022transformation,negredo2018fast}}, without accounting for optical losses based on Maxwell’s equations. Chips can have vastly dissimilar emission directions, consequently, the simulation domain for optimizing the PWB trajectory can be large. The primary challenge is to design this trajectory within seconds during the in-situ automated printing process, as the complex modeling required for large-scale PWB design is both computationally intensive and time-consuming. BPM is employed to model bent planar waveguides~\cite{lu2002beam}. However, it is not suitable for modeling 3D arbitrarily bent waveguides due to its reliance on the paraxial approximation and its inability to account for back-reflected waves. This method solves beam propagation in a waveguide by advancing the field stepwise along the longitudinal axis, under the assumption that propagation is predominantly directed along this axis with small transverse variations. This approach works well for simple straight waveguides but fails for complex, curved waveguides where the wave behavior deviates from simple beam evolution.

We adopt the following notation: Matrices and tensors we write with bold upright characters $\bm{\mathrm{M}}$. An inverse matrix is denoted as $\bm{\mathrm{M}}^{-1}$. The diagonal matrix $\bm{\mathrm{I}}$ is the identity matrix. If it is clear from the context, also vectors are written with bold upright characters $\bm{\mathrm{m}}$. In component form, we notate matrices and row vectors with brackets $[m_1,m_2,m_3]$, while column vectors are indicated by the transpose sign $[m_1,m_2,m_3]^T$. A special meaning is given to the vector mapping function $\bm{\mathrm{f}}(x,y,z)=[f_u(x,y,z),f_v(x,y,z),f_s(x,y,z)]^T$ and the reverse vector mapping function $\bm{\mathrm{f}}^{-1}(u,v,s)=[f_x^{-1}(u,v,s),f_y^{-1}(u,v,s),f_z^{-1}(u,v,s)]^T$, where $f^{-1}_{x,y,z}$ is the inverse function to $f_{x,y,z}$.

\subsection{TO method of modeling a freeform waveguide}
\begin{figure}[b!]
    \centering
    \includegraphics[width=\textwidth]{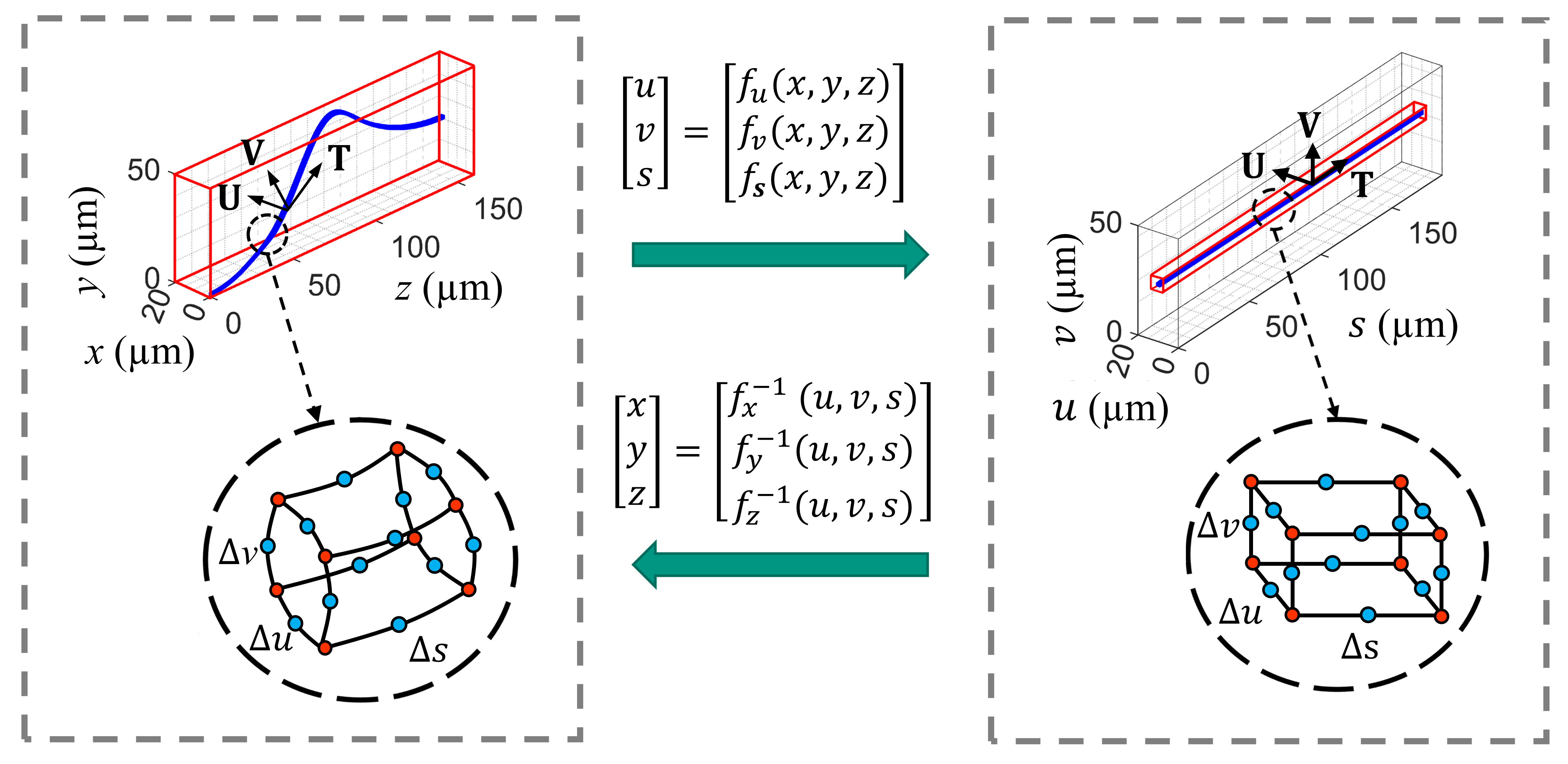}
    \caption{Second-order mapping \textbf{f} from the original curvilinear element to the virtual rectangular element, and reverse mapping $\textbf{f}^{\,-1}$. \revised{The red boxes in both spaces denote the simulation region enclosing the waveguide (blue).} The exemplary freeform waveguide in the real (\textit{x, y, z})-space has a simulation volume of $\qty{20} \times \qty{50} \times \qty{160}{\um}^3$, and the transformed waveguide in the virtual ($u$, $v$, $s$)-space has a simulation volume of $\qty{5} \times \qty{5} \times \qty{180}{\um}^3$ which is reduced by a factor of $35$. We employ a local right-handed trihedron at a point \((x_0, y_0, z_0)\) along the trajectory, defined by the unit vectors \((\mathbf{U}, \mathbf{V}, \mathbf{T})\), and we construct a  grid on the \((\mathbf{U}, \mathbf{V})\)-coordinate plane, which is normal to the trajectory direction \(\mathbf{T}\) in real space where the arc length is denoted by $s$. The left inset illustrates a hexahedral curvilinear mesh volume element in real space, which must be transformed into a hexahedral rectilinear mesh volume element in virtual space, as shown in the right inset. The curvilinear meshing facilitates a more accurate representation of the waveguide's contour. The red grid points represent the nodes to be used in electromagnetic numerical modeling, while the intermediate blue grid points serve as auxiliary points for a second-order mapping.}
    \label{fig:mapping}
\end{figure} 
The TO method~\cite{nesic2022transformation} greatly reduces the size of the simulation region for a 3D freeform waveguide in the (\textit{x, y, z})-space by transforming the curved waveguide into a straightened version within a virtual $(u, v, s)$-space, as shown by the red boxes in Fig.~\ref{fig:mapping}. Specifically, when mapping a domain from the (\textit{x, y, z})-coordinates to the virtual $(u, v, s)$-coordinates, only the material properties in the virtual domain need to be adjusted, while Maxwell's equations retain their original form\cite{nesic2022transformation}. The connection between the material properties in the real (\textit{x, y, z})-space and the virtual $(u, v, s)$-space is given by~\cite{nesic2022transformation, sun2017transformation}:
\begin{equation}
\begin{aligned} 
\bm{\mathrm{\epsilon}_r'}(u, v, s) = \frac{\bm{\mathrm{J}}(x,y,z)\cdot\bm{\mathrm{\epsilon}_r}(x,y,z)\cdot\bm{\mathrm{J}}^{T}(x,y,z)}{\det({\bm{\mathrm{J}}(x,y,z))}},\\
\bm{\mathrm{\mu}}_r'(u, v, s) = \frac{\bm{\mathrm{J}}(x,y,z)\cdot\bm{\mathrm{\mu}}_r(x,y,z)\cdot\bm{\mathrm{J}}^{T}(x,y,z)}{\det({\bm{\mathrm{J}}(x,y,z))}}.
\end{aligned}
\label{eqn:1}
\end{equation}
In these relations, the quantity $\bm{\mathrm{\epsilon}_r}$ denotes the permittivity tensors in real space, and $\bm{\mathrm{\epsilon}_r'}$ the tensor in virtual space. An equivalent relation holds for the permeability tensor $\bm{\mathrm{\mu}}_r$ in real and the tensor $\bm{\mathrm{\mu}}_r'$ in virtual space. The quantity \textbf{J} is the Jacobian matrix of the coordinate transformation function $[u, v, s]^T=\textbf{f}(x, y, z)$, and $\bm{\mathrm{J}}^{T}$ is its transpose,
\begin{equation}
\renewcommand{\arraystretch}{1.5} 
\mathbf{J} = \left[ \begin{array}{@{} c @{\hspace{3mm}} c @{\hspace{3mm}} c @{}}
\frac{\partial u}{\partial x} & \frac{\partial u}{\partial y} & \frac{\partial u}{\partial z} \\
\frac{\partial v}{\partial x} & \frac{\partial v}{\partial y} & \frac{\partial v}{\partial z} \\
\frac{\partial s}{\partial x} & \frac{\partial s}{\partial y} & \frac{\partial s}{\partial z} \\
\end{array} \right].
\label{eqn:2}
\end{equation}
The modulus of the determinant $\det({\bm{\mathrm{J}}(x,y,z)})$ measures the scale of the transformed volume element, which should not deviate too much in size from the real one. Otherwise, the tensor elements in virtual space could become too large or take unphysically small values, and so the employed algorithm would become inaccurate.

The transformed medium in virtual space is generally anisotropic~\cite{nesic2022transformation}, even if the structure in real space is made from isotropic material with permittivity tensor $\bm{\mathrm{\epsilon}_r}$ and permeability tensor $\bm{\mathrm{\mu}}_r$. Only if the trajectory lies in a plane, this method results in anisotropic, but diagonal permittivity and permeability tensors~\cite{nesic2022transformation}. To apply this method to non-planar waveguides, the numerical solver must be able to consider off-diagonal elements in the transformed material tensors. However, commercially available solvers cannot handle this situation~\cite{nesic2022transformation}. Another challenge arises when entering the transformed material properties in virtual space into a standard solver~\cite{nesic2022transformation}. \revised{Assigning the continuously varying, generally non-diagonal transformed tensors to the individual mesh cells of a general-purpose solver like Lumerical or CST on a fixed rectilinear grid is inefficient because of memory requirements and excessive computation time. Further, because of practical limitations set by the user interface, the material tensors cannot be assigned to each and every grid point, therefore the time-domain approach of~\cite{nesic2022transformation} approximates the medium by cuboid bricks with constant material. We avoid this problem by using a home-made MATLAB code, where we can easily define the transformed material properties at each grid point.}
\subsection{Curvilinear mapping}
For concentrating the numerical operations to the physically interesting space, we map the real space with Cartesian coordinates $(x,y,z)$ to a virtual-space with Cartesian coordinates $(u, v, s)$, Fig.~\ref{fig:mapping}. In real space, we employ a local right-handed trihedron at a point $(x_0,y_0,z_0)$ on the trajectory with unit vectors  \( (\mathbf{U}, \mathbf{V}, \mathbf{T}) \). Instead of the usual rectilinear volume element, we use a curvilinear volume element which approximates the geometry of the  waveguide  more closely  \cite{lu2002beam,viquerat20153d}. The local coordinates $(u, v, s)$ depend on the trihedron's origin $(x_0,y_0,z_0)$. To implement this mapping technique numerically, we need to define in the real space a hexahedral curvilinear mesh volume element $(\Delta u,\Delta v, \Delta s)$ (left inset Fig.~\ref{fig:mapping}), which needs to be transformed into a hexahedral rectilinear mesh volume element. This implies changing the $\bm{\mathrm{\epsilon}_r}$ and $\bm{\mathrm{\mu}_r}$ tensors according to Eq. (\ref{eqn:1}). To avoid unduly large deviations of the transformed tensors to the ones in real space, the Jacobian determinant should not deviate too much from one, and therefore the length of the edges in virtual space are chosen to have the same lengths as in real space (right inset Fig.~\ref{fig:mapping}). \revised{On the trajectory the mapping is an isometry, so \(\det(\mathbf{J}) = 1\). It departs from 1 only off-axis, where the real-space planes normal to the trajectory map onto equally spaced virtual planes. This stretches the inside of a bend and compresses the outside. The deviation grows at the borders of the computation region, i.e., farther away from the center line and for smaller bending radii. The mapping stays bijective only if these real-space planes do not intersect~\cite{nesic2022transformation}. The effect is shown in Fig.~\ref{fig:tobpm}(c).}

The planes perpendicular to the (tangential) trajectory direction \( \mathbf{T} \) are functions of the discrete trajectory points $(x_0,y_0,z_0)$~\cite{nesic2022transformation}, 
 \begin{equation}
\begin{aligned}
\left[ \begin{array}{@{}c@{}}
x(u, v, s) \\
y(u, v, s) \\
z(u, v, s) \\
\end{array} \right]
&= \left[ \begin{array}{@{}c@{}}
x_{0}(s) \\
y_{0}(s) \\
z_{0}(s) \\
\end{array} \right]
+u\bm{\mathrm{U}}(s) +v\bm{\mathrm{V}}(s).
\end{aligned}
\label{eqn:3}
\end{equation}

The \textit{\( s \)}-coordinate is measured in the direction of the unit tangent vector \( \mathbf{T} \) and  represents the arc length along the waveguide trajectory. Its starting point is defined as  \( \textit{\( s \)} = 0 \).  Equation~(\ref{eqn:3}) can be used to create a mesh grid on \((\mathbf{U}, \mathbf{V})\)-planes along the trajectory in real space. The procedure is illustrated in Fig.~\ref{fig:mapping}. Eq. (\ref{eqn:3}) leaves the freedom to select the orientation of the unit vectors \( \mathbf{U} \) and \( \mathbf{V} \) within the transverse \((\mathbf{U}, \mathbf{V})\)-coordinate planes at each point along the trajectory. To address this, we employ the rotation-minimizing frame (RMF) \cite{nesic2022transformation,bishop1975there}, which minimizes the rotational motion of the frame along the trajectory. The RMF is widely used in computer graphics and 3D modeling \cite{bechmann2003arbitrary}. For computing the RMF, we utilize a fast and simple approximation known as the double reflection method \cite{wang2008computation}. This approach requires knowledge of the trajectory's sample points, of the tangent vectors  \( \mathbf{T} \) at each sample point, and of the initial coordinate frame \( (\mathbf{U}_1, \mathbf{V}_1, \mathbf{T}_1) \) at the first sample point. The coordinate frames for the subsequent sample points along the trajectory are computed recursively \cite{nesic2022transformation,bechmann2003arbitrary}. The coordinate frame \( (\mathbf{U}_1, \mathbf{V}_1, \mathbf{T}_1) \) at the first sample point is selected such that \( \mathbf{T}_1 \) represents the tangent vector, with \( \mathbf{U}_1 \) and \( \mathbf{V}_1 \) mutually orthogonal in the plane perpendicular to \( \mathbf{T}_1 \). The discrete points on the \( \mathbf{UV} \)-plane are chosen to accurately represent the cross-sectional shape of a waveguide. 

Once the entire waveguide is meshed along its trajectory $[x_0,y_0,z_0]^{T}$, the resulting curvilinear hexahedral elements in each point of the trajectory are mapped to corresponding rectangular hexahedra in transformed $(u, v, s)$-space using a specific second-order function, which is uniquely defined for each volume element. It is generally more straightforward and intuitive to analytically represent the inverse \revised{vector} function \([x, y, z]^{T} = \textbf{f}^{\,-1}(u, v, s) \), which maps elements from the transformed space \( (u, v, s) \) back to the real space. Defining higher-order mappings requires additional grid points, blue dots in  Fig.~\ref{fig:mapping}. For a second-order mapping from virtual $(u, v, s)$-space to real (\textit{x, y, z})-space we write

\begin{equation}
    \revised{[x, y, z]^{T}} = \textbf{f}^{\,-1}(u, v, s)= \bm{a}_{1} + \bm{a}_{2}u + \bm{a}_{3}u^{2} + \bm{a}_{4}v+ \bm{a}_{5}v^{2} + \bm{a}_{6}s + \bm{a}_{7}s^{2}.
\label{eqn:4}
\end{equation}

\revised{Because the virtual $(u, v, s)$-grid is rectangular and the edges of the rectangular waveguide profile are aligned with the $u$, $v$ and $s$-axes, only one coordinate varies along an edge while the other two stay fixed. Therefore, Eq.~(\ref{eqn:4}) must be applied edge-wise. Since Eq.~(\ref{eqn:4}) is defined locally for each element, with the coordinates $(u, v, s)$ having their origin at the node, the $u$-edge takes the form $\left[ x(u,0,0),\, y(u,0,0),\, z(u,0,0) \right]^{\mathsf{T}} = \mathbf{f}^{-1}(u,0,0) = \mathbf{a}_{1} + \mathbf{a}_{2}\,u + \mathbf{a}_{3}\,u^{2}$ for determining the three vector coefficients $\mathbf{a}_{1\ldots3}$ fixed by the three points that define the edge (its two end nodes, red points in Figs.~\ref{fig:mapping} and~\ref{fig:unitcell}, and the intermediate auxiliary points, blue). A cross term such as $uv$, $us$, or $vs$ would require two coordinates to vary simultaneously along a single edge, which cannot happen according to our choice for a rectangular cell in virtual space. Because grid points in real $(x, y, z)$-space are placed according to Eq.~(\ref{eqn:3}) on planes normal to the true waveguide trajectory, and those planes rotate with the frame from one trajectory point to the next, the rotation and torsion of each frame is embedded in the coordinates of the grid points and are therefore taken care of by the inverse Jacobian $\mathbf{J}^{-1}$, whose three columns $\partial[x,y,z]^{\mathsf{T}}/\partial u$, $\partial[x,y,z]^{\mathsf{T}}/\partial v$ and $\partial[x,y,z]^{\mathsf{T}}/\partial s$ are obtained by differentiating the corresponding parabolic edge equations seen in Eq.~(\ref{eqn:4}).} The Jacobian matrix \( \mathbf{J} \) of the mapping \( \mathbf{f} \) is then obtained by taking the inverse of \( \mathbf{J}^{-1} \), \( \mathbf{J}=\left(\mathbf{J}^{-1}\right)^{-1} \). This matrix \( \mathbf{J} \) is then utilized to calculate the transformed permittivity and permeability tensors $\bm{\mathrm{\epsilon}_r'}$ and $\bm{\mathrm{\mu}}_r'$ as described in Eq. (\ref{eqn:1}). For further details refer to \ref{a:1}.
\subsection{TO-BPM Algorithm}
After transforming the waveguide into a straight configuration and having calculated the material properties at each grid point, we employ a full-vectorial finite-difference BPM to compute the electromagnetic fields. Since the transformed waveguide is straight, we  employ a paraxial BPM, which is computationally more efficient than a wide-angle version. \revised{In real space the propagation direction of a strongly bent waveguide varies along its trajectory, which translates into strongly varying material properties in virtual space. As a result, the slowly-varying envelope assumption is violated, and the magnetic tensor of the virtual waveguide deviates strongly from the identity matrix.} The BPM is a forward-propagating technique that ignores reflections. In waveguide simulations focused on transmission calculations, this limitation is  acceptable. A detailed description of this approach can be found in \cite{xu1994full}.

The BPM algorithm performs a stepwise propagation of an initial electric field vector through a volume, which is composed of \( N_{u} \times N_{v} \times N_{s} \) mesh cells with edge sizes \( \Delta u, \Delta v, \Delta s \) and varying tensors $\bm{\mathrm{\epsilon}_r'}$, $\bm{\mathrm{\mu}}_r'$ associated with the grid points. The \( s\)-axis is the direction of propagation in the virtual space.
We will demonstrate in Sect.~\ref{Results} that in all practical cases the relative permeability tensor elements in the main diagonal deviate by less than \qty{3}{\percent} from $1$, while its off-diagonal elements are smaller than $10^{-4}$. Therefore the material can be treated as non-magnetic with a permeability tensor $\bm{\mathrm{\mu_r}}=\bm{\mathrm{I}}$, and Maxwell's two first-order coupled differential equations reduce to two uncoupled second-order differential equations for $\bm{\mathrm{E}}$ (and $\bm{\mathrm{H}})$,
\begin{gather}
   \nabla \times \bm{\mathrm{E}} = -\operatorname j\omega\mu_0\bm{\mathrm{\mu}}_r \bm{\mathrm{H}} , \qquad
    \nabla \times \bm{\mathrm{H}} = \operatorname j\omega\epsilon_0\bm{\mathrm{\epsilon}_r} \bm{\mathrm{E}},\label{eqn:5a}\\
\nabla^2 \bm{\mathrm{E}} + k_0^2 \bm{\mathrm{\epsilon}_r} \bm{\mathrm{E}} = \nabla (\nabla \cdot \bm{\mathrm{E}}),\qquad k_0 =\left.\omega\right/c.
\label{eqn:5}
\end{gather}
The quantity $c$ is the speed of light in vacuum. For the transverse field vector $\bm{\mathrm{E}}_t(u,v,s)$ in the $u$-$v$ plane, we separate a rapidly varying spatial $s$-dependency $\operatorname{e}^{-\operatorname{j} n_0 k_0 s}$ from the amplitude vector $\bm{\mathrm{\Psi}}_t(u,v,s)$ which varies only slowly in $s$-direction.
\begin{equation}
\bm{\mathrm{E}}_t(u,v,s) = \bm{\mathrm{\Psi}}_t(u,v,s)\operatorname{e}^{-\operatorname{j} n_0 k_0 s}.
\label{eqn:6}
\end{equation}
where $n_0$ is a reference refractive index. Two cases can be taken: \( n_0 = n_{\text{eff}} \), and \( n_0 \neq n_{\text{eff}}\), where \( n_{\text{eff}} \) represents the effective index calculated as a weighted average value \cite{lu2023refractive},
\begin{equation}
n_{\text{eff}}^2 \approx \frac{\int\!\!\!\int n^2(u,v,s) |\bm{\mathrm{\Psi}}_t(u,v,s)|^2 \, du \, dv}{\int\!\!\!\int |\bm{\mathrm{\Psi}}_t(u,v,s)|^2 \, du \, dv}.
\label{eqn:neff}
\end{equation}
The separation of the rapidly varying part in  Eq.~\ref{eqn:6} allows the slowly varying envelope approximation (SVEA) \( \smash{\bigl|\partial^2 \bm{\mathrm{\Psi}}_t\big/\partial s^2\bigr|} \ll \smash{\operatorname 2n_0k \bigl|\partial \bm{\mathrm{\Psi}}_t\big/\partial s \bigr|} \), which is essential for the paraxial formulation of the BPM,
\begin{equation}
\operatorname j 2n_0 k_0 \frac{\partial}{\partial s} \bm{\mathrm{\Psi}}_t = \bm{\mathrm{P}}\bm{\mathrm{\Psi}}_t \quad \text{where} \quad \bm{\mathrm{\Psi}}_t=\left[ \begin{aligned} \mathit{\Psi_u} \\ \mathit{\Psi_v} \end{aligned} \right] ,\quad \bm{\mathrm{P}} = \left[ \begin{aligned} P_{uu} & P_{uv} \\ P_{vu} & P_{vv} \end{aligned} \right].
\label{eqn:7}
\end{equation}
The elements \( P_{ij} \) describe the components of the differential matrix operator $\bm{\mathrm{P}}$, see \ref{a:2} for further details. The finite-difference method can be used to solve Eq.~(\ref{eqn:7}). The weighing factor $0\le\alpha\le1$ determines to which extent the slices $l$ and $l+1$ contribute to the differential in Eq.~(\ref{eqn:7}),
\begin{align}
\frac{\partial}{\partial s} \bm{\mathrm{\Psi}}_t &= \frac{\bm{\mathrm{\Psi}}_t^{l+1} - \bm{\mathrm{\Psi}}_t^{l}}{\Delta s}, \\
\bm{\mathrm{P}} \bm{\mathrm{\Psi}}_t &= \bm{\mathrm{P}} \big(\alpha \bm{\mathrm{\Psi}}_t^{l+1} + (1-\alpha) \bm{\mathrm{\Psi}}_t^{l}\big) \, \overset{\alpha = 1/2}{=}  \bm{\mathrm{P}}\left(\frac{\bm{\mathrm{\Psi}}_t^{l+1} + \bm{\mathrm{\Psi}}_t^{l}}{2}\right),
\label{eqn:8a}\\
\underbrace{(2n_0 k_0\bm{\mathrm{I}}+\operatorname j \Delta s \,\alpha \bm{\mathrm{P}})}_{\bm{\mathrm{A}}} \underbrace{\bm{\mathrm{\Psi_t}}^{l+1}}_{\bm{\mathrm{X}}} &= \underbrace{[2n_0 k_0\bm{\mathrm{I}}-\operatorname j\Delta s (1-\alpha) \bm{\mathrm{P}}] \bm{\mathrm{\Psi_t}}^{l}}_{\bm{\mathrm{B}}}.
\label{eqn:8}
\end{align}
The slice index  \( l \) labels a plane normal to the propagation direction along the \( s \)-axis. This scheme is unconditionally stable for \( \alpha \geq 0.5 \), and power is conserved when \( \alpha = 0.5 \) as in Eq. (\ref{eqn:8a}) \cite{wang2008computation}. This method leads to a system of linear equations:
\begin{equation}
\bm{\mathrm{A}}\bm{\mathrm{X}} = \bm{\mathrm{B}}.
\label{eqn:9}
\end{equation}
where \( \bm{\mathrm{A}} \) is a sparse matrix, while the unknown $\bm{\mathrm{X}}$ and the known $\bm{\mathrm{B}}$ are vectors. MATLAB is used to efficiently solve this system of linear equations. Additionally, a perfectly matched layer (PML) is applied  at the boundaries of the computational domain \cite{shin2012choice}.
\section{Results and discussion} \label{Results}
\begin{figure}[htbp]
    \centering
    \includegraphics[width=0.94\textwidth]{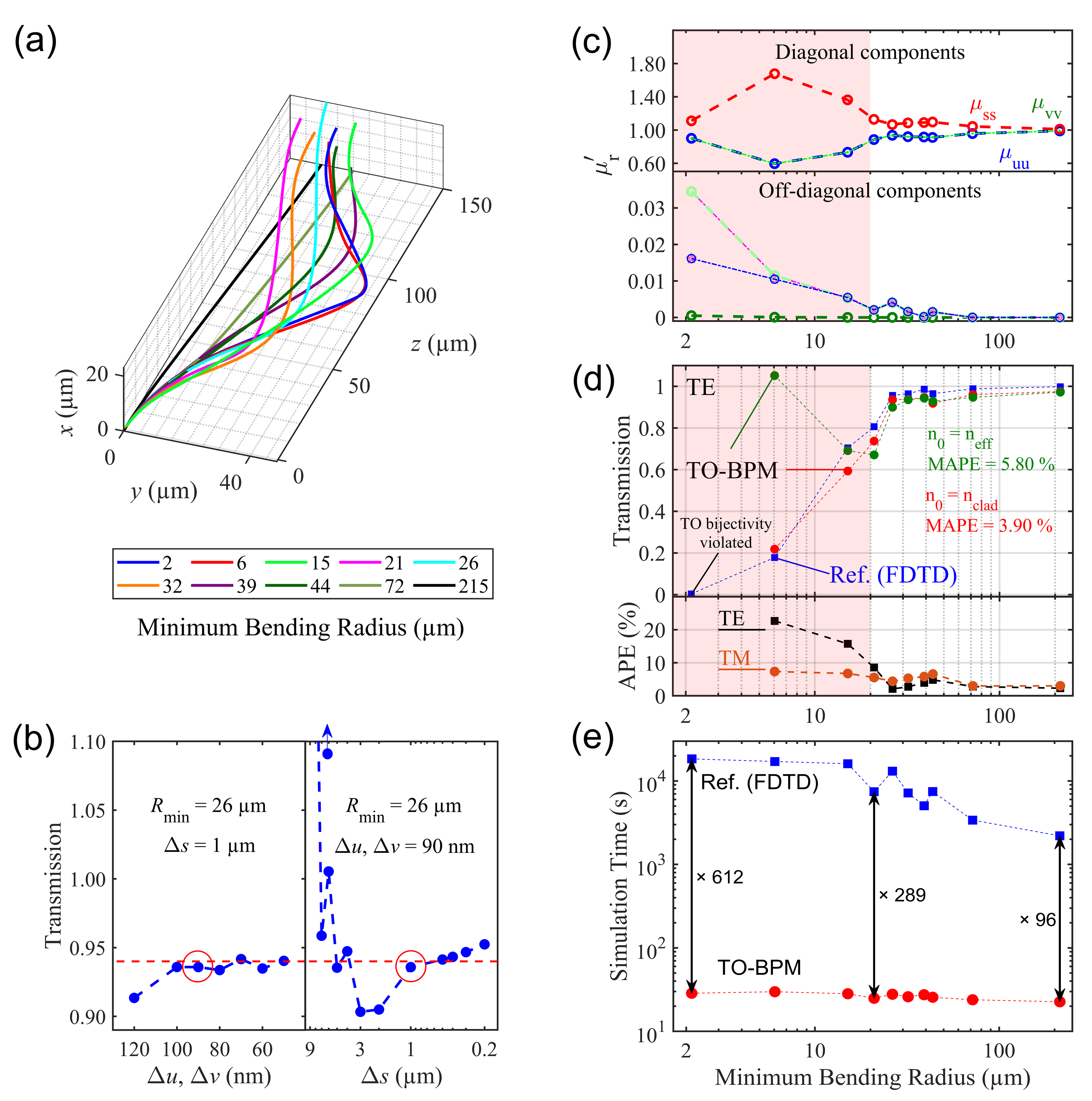}
    \caption{TO-BPM simulation results for exemplary non-planar bent PWBs. (a) Selection of exemplary PWBs with different minimum radii of curvature. Top: 3D display. Bottom: Cross-section in $yz$ and $xz$ plane. \revised{(b) Convergence test of TO-BPM algorithm for the waveguide with minimum bend radius $R_{\mathrm{min}}=\qty{26}{\um}$. PWB transmission with respect to $\Delta{{u}}$, $\Delta{{v}}$, when $\Delta{{s}} = \qty{1}{\um}$ (left), and with respect to $\Delta{{s}}$, when $\Delta{{u}}$, $\Delta{{v}} = \qty{90}{\nm}$ (right). The deviation from the FDTD reference (horizontal dashed line) is $\approx \qty{1}{\percent}$. Encircled points indicate the parameters used in this work. (c) Components of the transformed permeability tensor evaluated at the outermost computational grid point for different minimum bending radii of the investigated trajectories. The diagonal elements approach one, and the off-diagonal elements approach zero for bending radii larger than \qty{20}{\micro\meter}. (d) Top: Simulated power transmission at $\lambda = \qty{1.55}{\micro\meter}$ for the investigated PWBs, obtained with the TO-BPM algorithm (red/green circles) and reference FDTD simulation (blue squares), for transverse-electric (quasi-TE) input polarization. For $n_{\mathrm{o}} = n_{\mathrm{eff}}$ (green circles), the mean absolute percentage error (MAPE) is \qty{5.8}{\percent}, whereas for $n_{\mathrm{o}} = n_{\mathrm{clad}}$ (red circles), it reduces to \qty{3.9}{\percent} for $R_{\mathrm{min}} \geq \qty{20}{\micro\meter}$. White/light-red areas: accurate/inaccurate calculation regions. Bottom: Absolute percentage error (APE) of the TO-BPM transmission ($n_{\mathrm{o}} = n_{\mathrm{clad}}$) from the FDTD reference versus $R_{\mathrm{min}}$, for quasi-TE (black) and quasi-TM (orange) input polarization.} (e) Comparison of total simulation times for the TO-BPM algorithm (red curve) and reference FDTD simulations (blue curve), including the transformation time from the physical PWB to the straightened waveguide.}
    \label{fig:tobpm} 
\end{figure}
To test our TO-BPM algorithm, we apply it to analyze bent PWBs with trajectories that have minimum radii of curvature ranging from \( R_{\mathrm{min}} = \qty{2}{\um} \) to \( R_{\mathrm{min}} = \qty{215}{\um} \), as shown in Fig.~\ref{fig:tobpm}(a). These PWBs are chosen from a set of 3D randomly bent non-planar trajectories with smooth curvature variation. We have labeled them with their minimum radii of curvature. The rectangular cross-section of all the PWBs is $\qty{2}{\um} \times \qty{1.8}{\um}$ in the $x$ and $y$ directions, respectively. Core and cladding refractive indices are $n_{\mathrm{core}} = 1.53$ and $n_{\mathrm{clad}} = 1.36$, respectively. To determine the appropriate mesh cell sizes in our TO-BPM algorithm, we performed a convergence test. As illustrated in Fig.~\ref{fig:tobpm}(b) on the left, we varied $\Delta u$ and $\Delta v$ for a fixed mesh resolution of \( \Delta s = \qty{1}{\um} \), and  assessed the transmission of the PWB with \revised{\( R_{\mathrm{min}} = \qty{26}{\um} \) (left panel of Fig.~\ref{fig:tobpm}(b))}. The method was found to converge when $\Delta u , \Delta v = 90 \, \text{nm}$, with minimal oscillations of less than \qty{1}{\percent}. Subsequently, we fixed $\Delta u = \Delta v = 90 \, \text{nm}$ and varied $\Delta s$, as shown in Fig.~\ref{fig:tobpm}(b) on the right. The simulation results converged at \( \Delta s = \qty{1}{\um} \). The same behavior was observed for trajectories with higher minimum bending radius. Subsequently, we choose mesh sizes of $\Delta{{u}}=\Delta{{v}}=\qty{90}{\nm}$ in the directions perpendicular to the trajectory and a step size of $\Delta{\mathrm{s}}=\qty{1}{\um}$ in the direction along the trajectory. \revised{For trajectories at the very edge of the working region (\(R_{\mathrm{min}} \approx \qty{20}{\micro\meter}\)), the transmission does not converge monotonically with a refinement of the mesh, reflecting that we are near the accuracy limit of the TO-BPM. In this regime, the accuracy can be nonetheless improved by reducing the mesh size: With \(\Delta u = \Delta v = \qty{50}{\nano\meter}\) and \(\Delta s = \qty{500}{\nano\meter}\) the computed transmission approaches the FDTD reference. This refinement, however, raises the computation time to a few hundred seconds as compared to the standard mesh marked in Fig.~\ref{fig:tobpm}(b).}

In photonic integration, interconnects with offsets in the order of a few micrometer are typically required. For such scenarios, PWBs with \( R_{\mathrm{min}} \geq \qty{50}{\um} \) are sufficient to ensure smooth transitions and minimum optical losses. As shown in Fig.~\ref{fig:tobpm}(c), we examined the resulting permeability tensor $\bm{\mathrm{\mu}_r^\prime}$ from the TO approach and observed that for \( R_{\mathrm{min}} \geq \qty{20}{\um} \), the off-diagonal elements converge to zero, while the diagonal elements converge to one, leading to an identity matrix $\mathbf{I}$. This observation indicates that, within this range, our formulation Eqs.~(\ref{eqn:5a}),(\ref{eqn:5}) in which we assumed the material to be nonmagnetic, is reliable.

In Fig.~\ref{fig:tobpm}(d), we plot the simulated power transmission at $\lambda = \qty{1.55}{\um}$ obtained from the TO-BPM for each PWB trajectory characterized by $R_{\mathrm{min}}$, and compare the results to those obtained from a commercially available FDTD solver (Lumerical FDTD, Ansys Inc., Canonsburg, PA, USA)~\cite{ansyslumerical}. \revised{Each point in Fig.~\ref{fig:tobpm}(d) is a comparison of TO-BPM and FDTD for an identical waveguide design, which is relevant for the intended purpose of fabricating PWB. Each trajectory has a number of bends and a minimum bending radius. Because bends are the most prominent reason for loss, a transmission loss comparison for a selection of bending radii tells us the allowable accuracy region of operation: TO-BPM agrees with FDTD above $R_\mathrm{min} \approx \qty{20}{\um}$, and it departs below. The reason is that our TO-BPM assumes a non-magnetic medium, $\bm{\mu}'_r = \mathbf{I}$, (Eq.~(\ref{eqn:5})), whereas at tight bends the transformed permeability tensor deviates significantly from the identity tensor, $\bm{\mu}'_r \neq \mathbf{I}$, its off-diagonal elements becoming negligible only for $R_\mathrm{min} > \qty{20}{\um}$, Fig.~\ref{fig:tobpm}(c). Our method holds for both, transverse-electric (TE) and transverse-magnetic (TM) inputs (Fig.~\ref{fig:tobpm}(d)).}

At \( \alpha = 0.5 \), corresponding to the Crank-Nicolson scheme~\cite{xu1994unconditionally}, the method remains stable ensuring power conservation. Here, we investigate the effect of selecting different reference refractive indices \( n_0 \) for two cases in the relevant (white) region with  \( R_{\mathrm{min}} \geq \qty{20}{\um} \): \( n_0 = n_{\text{eff}} \), and \( n_0 = n_{\text{clad}} \), where \( n_{\text{eff}} \) represents the effective index \revised{of the straight waveguide in real space (\(n_{\text{clad}} = 1.36\), \(n_{\text{core}} = 1.53\), and with a rectangular cross-section as with the true PWB, namely $\qty{2}{\um} \times \qty{1.8}{\um}$ in the $x$ and $y$ directions, respectively)} calculated using Eq.~(\ref{eqn:neff}). By setting \( n_0 = n_{\text{eff}} \), the resulting MAPE is \revised{\qty{5.8}{\percent}}. On the other hand, setting \( n_0 = n_{\text{clad}} \), the smallest refractive index in the simulation region, reduces the MAPE to \revised{\qty{3.9}{\percent}}. This error represents the absolute deviation of the power transmission for the TO-BPM calculation relative to the FDTD reference values, averaged over all data points with \( R_{\mathrm{min}} \geq \qty{20}{\um} \). \revised{The choice $n_0 = n_\mathrm{clad}$ yields slightly closer agreement with FDTD for trajectories with $R_\mathrm{min} \geq \qty{20}{\um}$, but it gives much better results for tightly bent trajectories with $R_\mathrm{min} < \qty{20}{\um}$. In a bent waveguide, the mode becomes leaky and develops a significant component propagating in the cladding, so that the rapidly changing propagator is more appropriately referenced to by the cladding index $n_\mathrm{clad}$ than by the effective index $n_\mathrm{eff}$.}

\revised{The trajectories under investigation have three bends: One at the beginning and one at the ending of the waveguide, and one strong bend somewhere in-between. The beginning and ending bends are nearly identical because of their large bending radius. The trajectories are dissimilar in the sense that the middle bend is different not only in radius, but also in the 3D-direction of the bend for the various trajectories. It is therefore to be expected that the transmission factors do not follow a smooth curve.}

For better comparison, the TO-BPM and the FDTD simulations were run on the same system (AMD EPYC 7713P 64-Core 2\,GHz Processor, 512 GB DDR4-3200 Memory). As shown in Fig.~\ref{fig:tobpm}(e), our implementation is up to 600 times faster than conventional FDTD simulations for PWB bending radii smaller than $\qty{30}{\um}$, and 100 times faster for bending radii up to $\qty{200}{\um}$. With increasing bending radius, the resulting reduction in simulation volume for the FDTD calculation becomes less prominent, and the disparity in performance between the two methods diminishes. The computation times stay consistently below 30 seconds. The memory requirements of the TO-BPM (typically smaller than 2 GB) are significantly lower than that of the FDTD solver (typically larger than 100 GB).

\begin{figure}[htbp]
    \centering
    \includegraphics[width=\textwidth]{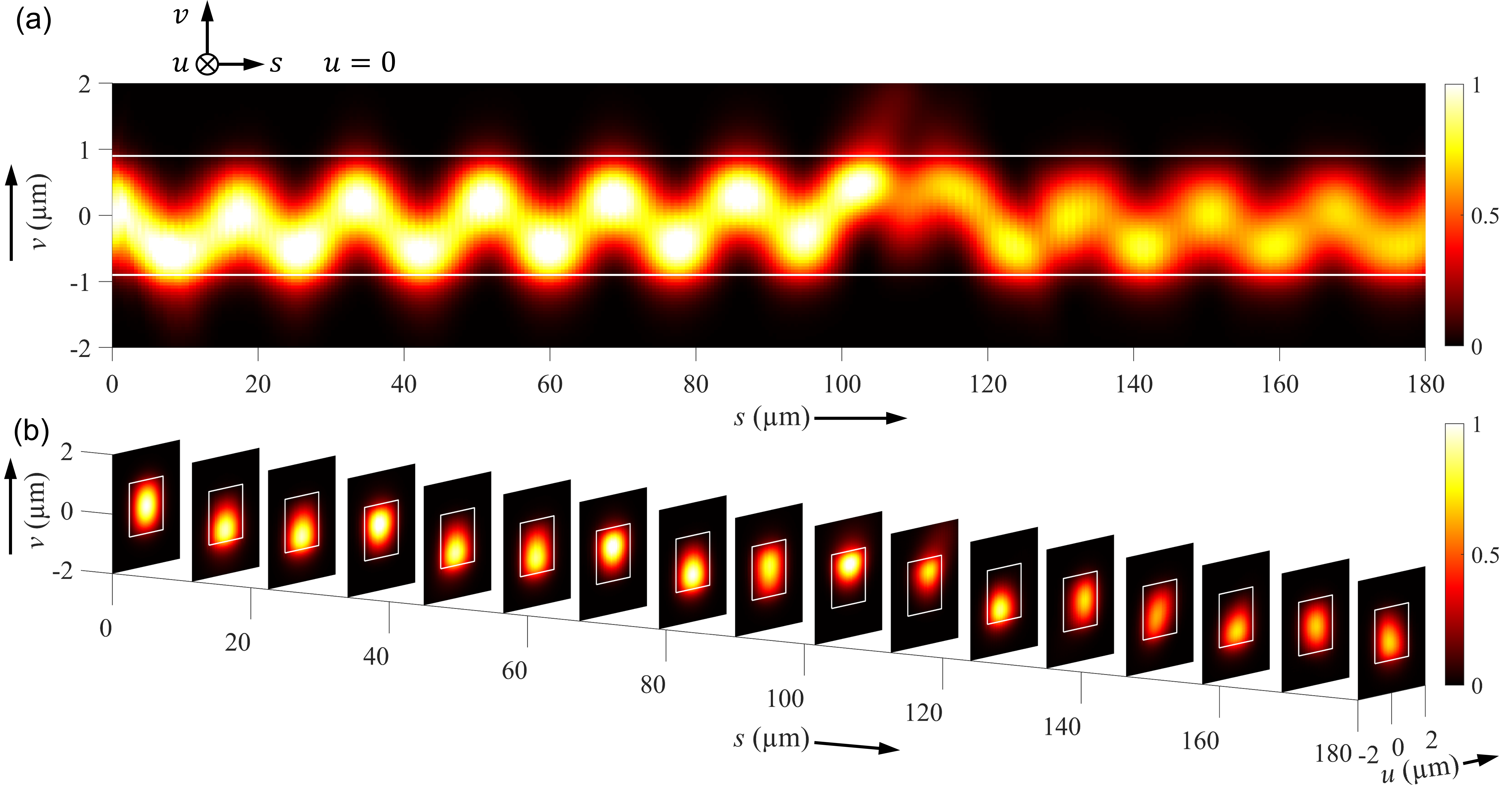}
    \caption{Simulated magnitude squared $|\bm{\mathrm{E}}|^2$ of the complex electric field vector at $\lambda = \qty{1.55}{\um}$ for the freeform waveguide with minimum bending radius $R_{\mathrm{min}}=\qty{15}{\um}$, \revised{significantly below the accuracy limit of \qty{20}{\micro\meter}.} (a) Field distribution in the $(u, v, s)$-space on the plane $u = 0$. (b) Field distribution in the $(u, v, s)$-space on \((\mathbf{U}, \mathbf{V})\)-planes normal to the propagation direction $s$.}
    \label{fig:fields}
\end{figure}

Figure~\ref{fig:fields} depicts the simulated electric field intensity $|\bm{\mathrm{E}}|^2$ for the PWB with a minimum bending radius of $R_{\mathrm{min}}=\qty{15}{\um}$, green curve in Fig.~\ref{fig:tobpm}(a). We demonstrate the results in the transformed $(u, v, s)$-space for TE polarized light, having a dominant electric field oriented along $x$ in $(x, y, z)$-space and along $u$ in $(u, v, s)$-space. Fig.~\ref{fig:fields}(a) illustrates the field distribution in $(u, v, s)$-space on the plane $u = 0$. Fig.~\ref{fig:fields}(b) presents the same field in a three-dimensional view, where successive \((\mathbf{U}, \mathbf{V})\)-planes normal to the propagation direction $s$ are shown along the waveguide trajectory. The bending in the $v$-direction is severe, with the smallest bend radius of $R_{\mathrm{min}}=\qty{15}{\um}$ at $s=\qty{110}{\um}$, which causes the mode to wiggle in the $v$-direction. This bend causes a noticeable leakage of light from the core into the cladding and subsequently into the perfectly matched layer surrounding the simulation region.

\revised{All calculations are performed in virtual \((u, v, s)\)-space. The fundamental mode (quasi-TE or quasi-TM) of the PWB is launched in the straight input section, and the transmission is the ratio of output to input power calculated from the Poynting vector at the two ports, without coupling to external waveguides. Since the input and output sections are straight, the Jacobian reduces to the identity matrix in these regions, so the physical and virtual descriptions coincide and the physical mode is launched in virtual space. The transmission in virtual space equals that in real space. If need arises, the physical fields can be calculated by \(\mathbf{E}(x, y, z) = \mathbf{J}^{\mathsf{T}} \mathbf{E}'(u, v, s)\) and \(\mathbf{H}(x, y, z) = \mathbf{J}^{\mathsf{T}} \mathbf{H}'(u, v, s)\). The field distributions in Fig.~\ref{fig:fields} are shown in virtual space because the real-space waveguide bends in three dimensions, and therefore the field profile is difficult to display.}
\section{Conclusion}
We introduced a computationally efficient TO-BPM algorithm that combines transformation optics (TO) with a full-vectorial BPM to model 3D-printed freeform PWBs. To validate the effectiveness of our approach, we conducted simulations of freeform PWBs with non-planar trajectories for a wide range of bending radii. Our method demonstrates a significant improvement in computational efficiency, achieving a reduction of computation time by a factor of up to 600 compared to a reference FDTD simulations in real 3D space. The TO-BPM method achieved a mean absolute percentage error of \revised{\qty{3.9}{\percent}} compared to the FDTD method. Notably, this work represents, to the best of our knowledge, the first successful combination of transformation optics with  a fully vectorial BPM, which is capable of handling 3D freeform waveguides with non-planar trajectories. We believe that the TO-BPM technique has the potential to facilitate the design and optimization of 3D freeform photonic wire bonds.
\appendix 
\renewcommand{\thesection}{Appendix \Alph{section}}
\section{Second-order mapping of curvilinear mesh elements} \label{a:1}
\begin{figure}[b!]
    \centering
    \includegraphics[width=\textwidth]{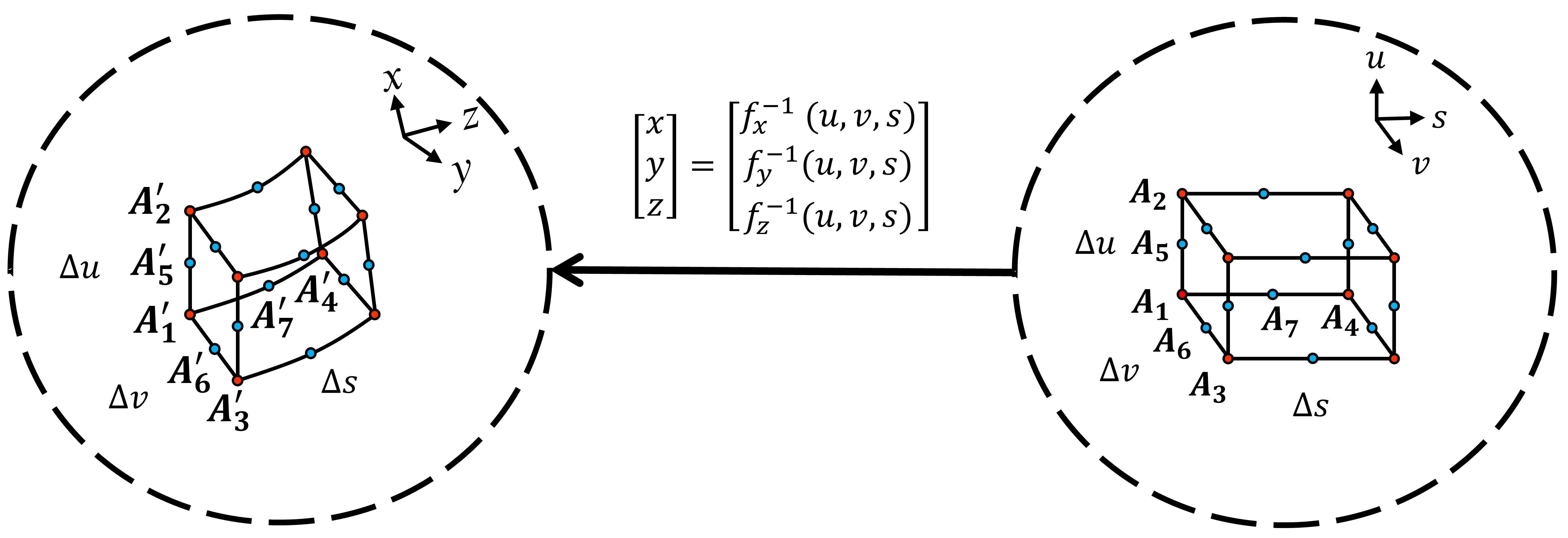}
    \caption{Second-order mapping from the virtual element to the original curvilinear element. The lateral mesh sizes are $\Delta u=\Delta v=\qty{90}{\nm}$, and the longitudinal step size amounts to \revised{$\Delta s=\qty{1}{\micro\meter}$} in both spaces.}
    \label{fig:unitcell}
\end{figure}
This section describes the procedure for calculating the coefficients of the second-order mapping \(\bm{a}_{k}\) in Eq. (\ref{eqn:4}). In this work, we assume a rectangular cross-section for our PWBs, which results in straight edges of the mesh cells along the \( u \) and \( v \) directions, as illustrated in Fig.~\ref{fig:unitcell} and Fig.~\ref{fig:mapping}. Henceforth, the edge linking the vertices \(\bm{A'_1}\) and \(\bm{A'_2}\) via \(\bm{A'_5}\) as shown in Fig.~\ref{fig:unitcell}, will be represented as \(A'_1-A'_2\). To compute the coefficients \(a_{k}\) based on the vertex coordinates, each edge is parameterized by selecting a suitable coordinate system \((u, v, s)\) within the second-order mapping expression. For instance, for the edge \(A'_1 - A'_2\), a suitable parameterization is given by
\begin{equation}
\begin{bmatrix}
A'_x \\ 
A'_y \\ 
A'_z
\end{bmatrix}
= \textbf{f}^{\,-1}(u, 0, 0) =
\begin{bmatrix}
a_{1x} \\ 
a_{1y} \\ 
a_{1z}
\end{bmatrix}
+
\begin{bmatrix}
a_{2x} \\ 
a_{2y} \\ 
a_{2z}
\end{bmatrix} \, u
+
\begin{bmatrix}
a_{3x} \\ 
a_{3y} \\ 
a_{3z}
\end{bmatrix} \, u^2.
\label{eqn:10}
\end{equation}
Next, this parameterization is assessed at \( u = 0 \), \( u = \frac{1}{2} \Delta u\), and \( u = \Delta u \), which corresponds to \( A'_1 \), \( A'_5 \), and \( A'_2 \), respectively. Consequently, this results in,
\begin{equation}
\begin{aligned}
\bm{a_1} &= \bm{A'_1}, \\
\bm{a_1} + \bm{a_2} \frac{\Delta u}{2} + \bm{a_3}  \left(\frac{\Delta u}{2}\right)^2 &= \bm{A'_5}, \\
\bm{a_1} + \bm{a_2} \Delta u + \bm{a_3} \Delta u^2 &= \bm{A'_2}.
\end{aligned}
\label{eqn:11}
\end{equation}
By applying the same procedure to the other edges, \(A'_1-A'_3\) and \(A'_1-A'_4\), with modified parameterization, the resulting second-order mapping coefficients are as follows,
\begin{equation}
\begin{aligned}
\bm{a_1} &= \bm{A'_1}, \\
\bm{a_2} &= \frac{-3\bm{A'_1} - \bm{A'_2} + 4\bm{A'_5}}{\Delta u}, \\
\bm{a_3} &= \frac{2\bm{A'_1} + 2\bm{A'_2} - 4\bm{A'}_5}{(\Delta u)^2}, \\
\bm{a_4} &= \frac{-3\bm{A'_1} - \bm{A'_3} + 4\bm{A'_6}}{\Delta v}, \\ 
\bm{a_5} &= \frac{2\bm{A'_1} + 2\bm{A'_3} - 4\bm{A'_6}}{(\Delta v)^2}, \\
\bm{a_6} &= \frac{-3\bm{A'_1} - \bm{A'_4} + 4\bm{A'_7}}{\Delta s}, \\
\bm{a_7} &= \frac{2\bm{A'_1} + 2\bm{A'_4} - 4\bm{A'_7}}{(\Delta s)^2}.
\end{aligned}
\label{eqn:12}
\end{equation}
\revised{The three-point fit along each of the three edges emanating from a node yields the per-edge quadratic coefficients, which determine the elements of the inverse Jacobian $\mathbf{J}^{-1}$ at that node. Continuity between neighboring edges holds, since any shared edge is defined by the same beginning, ending and auxiliary points, with no gaps or overlaps.} By applying this procedure at each grid point (denoted as red points in Fig.~\ref{fig:unitcell}), the components of the \( \mathbf{J}^{-1} \) matrix can be obtained, along with the transformed permittivity (\( \bm{\mathrm{\epsilon}_r'} \)) and permeability (\( \bm{\mathrm{\mu}}_r' \)) tensors.
\renewcommand{\thesection}{Appendix \Alph{section}}
\section{Anisotropic full-vectorial BPM} \label{a:2}
For an anisotropic and non-magnetic material, the dielectric tensor takes the form of,
\begin{equation}
\bm{\mathrm{\epsilon}_r} = \begin{pmatrix}
n_{uu}^2 & n_{uv}^2 & n_{us}^2 \\
n_{vu}^2 & n_{vv}^2 & n_{vs}^2 \\
n_{su}^2 & n_{sv}^2 & n_{ss}^2
\end{pmatrix}.
\label{eqn:17}
\end{equation}
The differential operators in Eq. (\ref{eqn:8}) are derived by substituting Eq. (\ref{eqn:7}) into the vector wave equation and applying the slowly varying envelope approximation \cite{xu1994full},
\begin{align}
P_{uu} \mathit{\Psi}_u &= \frac{\partial^2 \mathit{\Psi}_u}{\partial v^2} 
+ \frac{\partial}{\partial u} \left[ \frac{1}{n_{ss}^2} \frac{\partial}{\partial u} \left(n_{uu}^2 \mathit{\Psi}_u \right) \right]  
+ \frac{\partial}{\partial u} \left[ \frac{1}{n_{ss}^2} \frac{\partial}{\partial v} \left(n_{vu}^2 \mathit{\Psi}_u \right) \right] 
+ \left(n_{uu}^2 - n_0^2 \right) k_0^2 \mathit{\Psi}_u, \label{eqn:18} \\
P_{vv} \mathit{\Psi}_v &= \frac{\partial^2 \mathit{\Psi}_v}{\partial u^2} 
+ \frac{\partial}{\partial v} \left[ \frac{1}{n_{ss}^2} \frac{\partial}{\partial v} \left(n_{vv}^2 \mathit{\Psi}_v \right) \right]  
+ \frac{\partial}{\partial v} \left[ \frac{1}{n_{ss}^2} \frac{\partial}{\partial u} \left(n_{uv}^2 \mathit{\Psi}_v \right) \right]  
+ \left(n_{vv}^2 - n_0^2 \right) k_0^2 \mathit{\Psi}_v, \label{eqn:19} \\
P_{uv} \mathit{\Psi}_v &= n_{uv}^2 k_0^2 \mathit{\Psi}_v 
+ \frac{\partial}{\partial u} \left[ \frac{1}{n_{ss}^2} \frac{\partial}{\partial v} \left(n_{vv}^2 \mathit{\Psi}_v \right) \right] 
+ \frac{\partial}{\partial u} \left[ \frac{1}{n_{ss}^2} \frac{\partial}{\partial u} \left(n_{uv}^2 \mathit{\Psi}_v \right) \right] 
- \frac{\partial^2 \mathit{\Psi}_v}{\partial u \partial v}, \label{eqn:20} \\
P_{vu} \mathit{\Psi}_u &= n_{uv}^2 k_0^2 \mathit{\Psi}_u 
+ \frac{\partial}{\partial v} \left[ \frac{1}{n_{ss}^2} \frac{\partial}{\partial u} \left(n_{uu}^2 \mathit{\Psi}_u \right) \right] 
+ \frac{\partial}{\partial v} \left[ \frac{1}{n_{ss}^2} \frac{\partial}{\partial v} \left(n_{vu}^2 \mathit{\Psi}_u \right) \right]  
- \frac{\partial^2 \mathit{\Psi}_u}{\partial v \partial u}. \label{eqn:21}
\end{align}
\section*{Funding}
Deutsche Forschungsgemeinschaft (EXC-2082/1 - 390761711, SFB 1173); European Research Council (773248); GOSPEL (403187440); ATHENS (101167540); Karlsruhe School of Optics \& Photonics (KSOP).

\section*{Acknowledgements}
This research has been funded by the Deutsche Forschungsgemeinschaft (DFG, German Research Foundation) under Germany's Excellence Strategy via the Excellence Cluster 3D Matter Made to Order (EXC-2082/1 - 390761711) and via the DFG Collaborative Research Center (CRC) WavePhenomena (SFB 1173), in part by the European Research Council via the ERC Consolidator Grant TeraSHAPE 773248, by GOSPEL under Grant 403187440, by ATHENS under Grant 101167540, and by the Karlsruhe School of Optics \& Photonics (KSOP).

\section*{Disclosures}
The authors declare no conflicts of interest.

\section*{Data Availability Statement}
Data underlying the results presented in this paper are not publicly available at this time but may be obtained from the authors upon reasonable request.


\bibliography{sample}

\end{document}